# *SYNTAX: A COMPUTER PROGRAM TO COMPRESS A SEQUENCE AND TO ESTIMATE ITS INFORMATION CONTENT*

Miguel A. Jiménez-Montaño*,   Werner Ebeling[+],   Thorsten Pöschel[++]

* Departamento de Física y Matemáticas, Universidad de las Américas/Puebla[1].
  Sta. Catarina Mártir, 72820 Puebla, México
+ Institut für Theoretische Physik, Hunboldt-Universität,
  Invalidenstr. 42, D-O 1040 Berlin, Germany
++ Laboratorie de Physique Mecanique des Milieux Heterogenes, Ecole Superieure de
   Physique et Chimie Industrielle, 10 rue Vauquelin, 75231 Paris Cedex 05 France

**Abstract**

The determination of block-entropies is a well established method for the investigation of discrete data, also called symbols (7). There is a large variety of such symbolic sequences, ranging from texts written in natural languages, computer programs, neural spike trains, and biosequences. In this paper a new algorithm to construct a short context-free grammar (also called program or description) that generates a given sequence is introduced. It follows the general lines of a former algorithm, employed to compress biosequences (1,2) and to estimate the complexity of neural spike trains (4), which uses as valuation function the , so called , grammar complexity (2). The new algorithm employs the (observed) block-entropies instead. A variant, which employs a corrected "observed entropy", as discussed in (7) is also described. To illustrate its usefulness, applications of the program to the syntactic analysis of a sample biological sequences (DNA and RNA) is presented.

## *Objective*

In this communication we present a new algorithm to compress a sequence of symbols. Although the algorithm can be applied to sequences in any alphabet, for the sake of simplicity, we are going to illustrate its performance by applying it to binary sequences. As a side result, an estimation of the *information content* of the given sequence is obtained.
- To show its usefulness, some simple examples are discussed as well as its possible application to the syntactic analysis of biosequences.

## *Previous Algorithm*

In former papers (1,2,3) an algorithm to compress sequences was described and applied to the description of biosequences (4). Independently, in (5) a similar algorithm was introduced and applied to the discovery of phrase structure in natural language. Briefly, our former procedure is as follows:

---

[1] This work was supported by CONACyT (México) , Project: 1932-E9211.

All the subwords of length two are formed to make, with each one, a search over the whole string to determine the most frequent one. The most frequent pattern is substituted by a non-terminal symbol (syntactic category) in all its appearances in the sequence under analysis, with the condition that its frequency is greater than two. This operation is repeated until there are no more strings of length two which occur more than two times. Then one searches for strings of length equal or greater than three that are repeated at least two times substituting, after the search, the longest one by a non-terminal symbol. In this way a context-free grammar which generates the original sequence is obtained.

The above algorithm was not intended for sequence compression *per se*, but was designed to estimate the sequence *grammar complexity* (1). This quantity is defined as follows:

Let G be a context-free grammar with alphabet
$V = V_T \cup V_N$ which generates only the word **w** ( if **s** -> **w₁**
and s -> **w**$_2$ , with **w₁** and **w**$_2$ ε $V_T$ then **w**$_1$ = **w**$_2$ ). These grammars are called "programs" or "descriptions" of the word **w**. The above described algorithm was not intended for sequence compression *per se*, but was designed to estimate the sequence *grammar complexity* (1). This quantity is defined as follows:

The complexity of a production-rule A -> q is defined by an estimation of the complexity of the word in the right-hand side: $q \to a_1^{v_1} \ldots a_m^{v_m}$ :

$$K(A \to q) = \sum_{j=1}^{m} \{ [\log_2 v_j] + 1 \},$$

where $a_j \in V_T \cup V_N$, for all $j = 1, \ldots, m$; and $[x]$ denotes the integral part of a real number.

The complexity K(G) of a grammar G is obtained by adding the complexities of the individual rules. Finally, the estimation of the complexity of the original sequence is:

$$K(G(\mathbf{w})) = \min \{K(G) \mid G \to \mathbf{w}\}.$$

*The New Algorithm*

The problem with the above definition of complexity is that the terminals (letters of the alphabet) and the non-terminals (syntactic categories) are treated on the same footing. Since the *self-information (see below)* of letters and categories is different, this fact should be taken into account in the calculation of the complexity.

Before introducing the new algorithm we need to introduce some preliminary concepts. The well known Shannon entropy is:

$$H_1 = \sum_{j=1}^{\lambda} - p_j \log_2 p_j$$

where $\lambda$ is the number of letters in the alphabet.

A straightforward generalization are the block-entropies:

$$H_n = \sum_j - p_j^{(n)} \log_2 p_j^{(n)}$$

Here, the summation is to be carried out over all n-words (n-tuples of letters) with non-vanishing probability $p_j^{(n)}$. $H_n$ measures the *information content* of an n-word, i.e. the mean number of binary questions to guess an n-word generated by the underlying process. $H_n$ becomes maximal if all symbols are equidistributed and stistically independent. For independent letters any letter requires $H_1$ binary questions and, hence, $H_n = n \cdot H_1$. Consequently, the differences $H_1 - H_n / n$ can be regarded as a measure of correlations between symbols.

For stationary and ergodic sources the entropy of the source: $h = \lim_{n \to \infty} H_n / n$

exists and gives the information per letter by taking into account all statistical dependencies. In a sense, $H_n$ resp.
$h_n = H_{n+1} - H_n$ are ideal candidates to detect structures in symbolic sequences since they respond to any deviations from statistical independence. However, their estimation from finite samples appears to be problematic due to the combinatorial explosion: the number of possible n-words grows like $\lambda^n$ which reaches astronomical numbers even for moderate word lengths n. (see 6, for example).

For our present problem the situation is even worse since we do not know the source that generated the given sequence, all we know is the sequence itself. Thus, the approach starts by stimating the entropy by means of the "observed entropy":

$$H^{obs} = \sum_j - k_j / N \log_2 k_j / N \quad,$$

where the $k_j$ denote the number of occurrences of a certain word. For our algorithm we are going to need words of length one (letters) and two (pairs). Thus, we shall call the expression $I = \log_2 k_j / N$ the (observed) *self-information* *per* letter (resp. pair). For binary sequences we have for the letters : $k_1 = $ # of ones and $k_0 = $ # of zeros, and N is the length of the sequence. And, for a given pair, k is the number of occurrences of the pair, and N the number of pairs in the sequence (counted overlapping). In (6) an improved estimation of $H^{obs}$ is discussed :

$$H = H^{obs} + \sum_j \frac{1}{2 N \ln 2} \quad.$$

However, for short words this correction may be neglected (see 7).

Description of the *new algorithm*:

Let S -> q be the trivial grammar that generates the sequence q. The complexity of q, as estimated from this gramnmar, is simply  C = N * $H^{obs}$ .

Now, recalling our former algorithm, one looks for the most frequent pair defines the first non-terminal and calculates its self-information . For example, for the pair 10, one introduces the production-rule: $s_1$ -> 10  with the corresponding self-information :
$I_{10}$ = $\log_2$ $k_{10}/N_{10}$ . Then the pair $s_1$ is substituted in sequence q and the process repeated, as explained before.

The complexity of a rule is the summation of the self-information of the symbols occurring in its right-hand side:

$$C(A \rightarrow \mathbf{w}) = \sum_j I_j$$

The complexity C(G) of a grammar G is obtained by adding the complexities of the individual rules. As more rules are introduced C(G) diminishes up to a point , then the introduction of new rules increases the complexity again and the process must stop. The estimation of the complexity of the original sequences is:

$$C(G\{q\}) = \min \{ C(G) \mid G \rightarrow q \}.$$

This quantity is an estimation of the *information content* of the sequence, which takes into account its block structure.
The above algorithm may be best understood with the help of some simple

*EXAMPLES*

**1**. Lets consider the random sequence:
 V = 01001110100111101000001100101101.
This sequence may be generated by the grammar $G_0$ :

S -> $S_0$ 0 $S_0$ 11 $S_0$ 0 $S_0$ 111 $S_0$ 0000 $S_0$ 10 $S_0$ $S_0$ 1 $S_0$
$S_0$ -> 01.

We calculate the self-information from:   p(1) = p(0) = 16/32 = 0.5 ,then  $I_1 = I_0 = 1$.
And from:
p(01) = 9/31, then $I_{01}$ = - $\log_2$ (9/31) = 1.78427 bits.

Therefore,  C($S_0$ ->01) = 1 + 1 = 2
 C(S ->q) = 9 x $I_{01}$ + 14 = 9 x 1.78427 + 14 =   30.05843,
Thus, C($G_0$) =  2 + 30.05843 = 32.05843 > 32 bits! .Which is greater than the complexity of V as estimated from the trivial grammar (the sequence itself): C (S-> V) = N * $H_1$ =32 bits .

If the process is continued by adding the rule:
$S_1 \to 0\ S_0$   one gets the grammar $G_1$:

$S \to S_0\ S_1\ 11\ S_0\ S_1\ 111\ S_0\ 000\ S_1\ 1\ S_1\ S_0 1\ S_0$
$S_1 \to 0\ S_0$
$S_0 \to 01.$

with $C(G_1) = 30.54322$, which is even worst.

The process may be continued with the addition of the rules $S_2 \to S_1\ 1$ and
$S_3 \to 1\ S_0$, producing the grammar $G_4$:

$S_4 \to S_0\ S_2\ S_3\ S_2\ 1\ S_3 000\ S_2\ S_1\ S_0\ S_3$
$S_3 \to 1$
$S_2 \to S_1\ 1$
$S_1 \to 0\ S_0$
$S_0 \to 01.$

which has $C(G_4) = 32.4541$ bits. However, according to the old algorithm, the grammar complexity of V, as estimated from the sequence itself is:
$K(S \to V) = 28$, and estimated with $G_4$ is $K(G_4) = 20$, having achieved a reduction of a random sequence which is a contradiction according to algorithmic information theory. This problem is avoided with the new algorithm.

**2**. Lets consider now a regular sequence:
W -> 11110000111100001111000011110000
Complexity Init: $H_1 * N = 32.0000$; Length $N = 32$

It can be generated with the grammar G:

$S \to S_0\ S_0\ S_0\ S_0$
$S_0 \to 11110000$

obtained with the new algorithm after removing redundat rules, which occur les than three times in the final grammar ( See **Fig. 1**). $I_0 = I_1 = 1$, $I_{s0} = I(11110000) = 0.80735$.
Therefore, $C = 4 \times 0.80735 + 8 = 11.2294$ bits $< 32.0000$ bits.
A substancial reduction is obtained, due to the periodicity of the sequence. In this way the sequence is compressed and its information content estimated.

Further examples and results are displayed in **Table 1.**
In Fig. 3 the result of applying the algorithm to a fragment of DNA is displayed

# CONCLUSIONS

The algorithm presented in this communication represents a substancial improvement over our former algorithm. Although the grammar complexity has been applied successfully to the analysis of biosequences and neural spike trains, we expect that the former results will be improved with the application of the new algorithm. The generality of the algorithm permits its application in other fields as well. For example to detect phrase structure of natural languages and for sequence compression in general.

**Acknowledgements**
The assistance of Miguel Angel Hernández M. in writing the code and the final presentation is warmly recognized. This work was supported by CONACyT (mexico), Project: 1932-E9211.

Sequence: W
11110000111100001111000011110000
Complexity Init:   32.0000  Length: 32

Iter: 1
 Length: 24
 Complexity:  33.6336
 1   1   1   1   S0   S0   1   1   1   1   S0   S0   1   1   1   1   S0   S0   1   1   1
 1   S0   S0
 S0 -> ( 0  0)  freq: 8, info: 1.95420

Iter: 2
 Length: 16
 Complexity:  31.8221
 S1   S1   S0   S0   S1   S1   S0   S0   S1   S1   S0   S0   S1   S1   S0   S0
S0 -> ( 0  0)  freq: 8, info: 1.95420
S1 --> ( 1  1) freq: 8, info: 1.52356

Iter: 3
Length: 12
 Complexity:  27.7245
 S0 -> ( 0  0)  freq: 8, info: 1.95420
 S1   S1   S2   S1   S1   S2   S1   S1   S2   S1   S1   S2
 S1 --> ( 1  1) freq: 8, info: 1.52356
 S2 --> ( S0  S0)  freq: 4, info: 1.90689

Iter: 4
 Length: 8
 Complexity:  23.2708
 S1   S3   S1   S3   S1   S3   S1   S3
 S0 -> ( 0  0)  freq: 8, info: 1.95420
 S1 --> ( 1  1) freq: 8, info: 1.52356
 S2 --> ( S0  S0)  freq: 4, info: 1.90689
 S3 --> ( S1  S2)   freq: 4, info: 1.45943

Iter: 5
 Length: 4
 Complexity:  17.5513
 S4   S4   S4   S4
 S0 -> ( 0  0)  freq: 8, info: 1.95420
 S1 --> ( 1  1) freq: 8, info: 1.52356
 S2 --> ( S0  S0)  freq: 4, info: 1.90689
 S3 --> ( S1  S2)   freq: 4, info: 1.45943
 S4 --> ( S1  S3)   freq: 4, info: 0.807355

Iter: 6, Erase Rules

Result (minimum complexity):
 Length:  4
 Complexity:    11.2294
 S4   S4   S4   S4
Rules:
0 --> ( 0)  info:    1.00000
1 --> ( 1)  info:    1.00000
S4 --> (  1   1   1   1   0   0   0   0)  freq: 4,  info:    0.807355

**Implementation**

The program is implemented in Fortran77 on a Unix plataform. The test cases were executed in a Sun SPARCstation IPC with 24 Mb of RAM and running SunOS R 4.1.3. The program searches for a master file that lists the sequences to be analyzed, this list was made in order to make the program easier to use. Each sequence must be stored in a text file with the length of the sequence in the first row and each subsequent symbol in a row. For reasons concerning to the implementation of the program, the symbols must be positive integers.

The Algorithm

The program reads the sequence and computes its complexity in the terms explained above. The first measure of complexity is equivalent to the Shanon's entropy multiplied by the length of the sequence. Then, the algorithm searches for the most frequent pair and, if such frequence happens to be greater than a predefined freq_accept, defines a rule(i) for that pair. Next, that rule is replaced on the original sequence, computing the complexity of the sequence again. This procedure is iterated until the frequence of all pairs is less than the freq_accept. Afterwards, the program erases the rules that appear less times than the freq_accept, counting on both the compressed sequence and the rules generated.

The program saves the complexity obtained in each iteration and after erasing each of the rules. This way, the complexity of the sequence is the minimum of all the computed results.

**Tabulation of the complexity for several examples**

| key after | Sequence | Complexity | Complexity supressing redundant rules |
|---|---|---|---|
| X | 00000000000000000000000000000000 | 0.0000 * | |
| 13 | 10101010101010101010101010101010 | 8.95159 | |
| Y | 10011001100110011001100110011001 | 12.5210 | 11.2294 |
| W | 11110000111100001111000011110000 | 17.5513 | 11.2294 |
| 10 | 00110011110011001111111100000000 | 29.0164 | |
| U | 00001001100000010100000000100000 | 22.2788 * | |
| 8 | 10010111000011011001011100001101 | 32.0000 * | |
| 3 | 10101011011001100111001110011111 | 28.2829 | 28.2770 |
| 11 | 11011001100011000011000001100000 | 25.9807 | 22.3803 |
| 5 | 01001010000000001111111110100 11 | 32.0000 * | |
| 7 | 01011011000000001111111100000011 | 32.0000 * | |
| V | 01001110100111101000001100101101 | 32.0000 * | |
| 1 | 10001001101100111011010100110101 | 30.1816 | |
| 4 | 11101111111010100010101100000000 | 32.0000 * | |
| 9 | 10011100000011011011000000111001 | 31.6384 * | |
| 6 | 00010010011000011110110110011110 | 32.0000 * | |
| 2 | 01011010111100000010100100111000 | 31.6384 * | |

*(\*) uncompressed sequences.*

*Note: The sequences and the keys are from reference ( 3 ) .*

Table 1

*Pseudocode*

```
read from 'syntax.sequences', number of sequences
for s=1 to number of sequences
  read sequence(s)
  i= 1
  freq= frequence of most frequent pair
 while (freq >= freq_accept) do
   rule(i)= most frequent pair
   replace rule(i) in sequence(s)
   if ( cpx <= cp_min) then
      cpx_min= cpx
   freq= frequence of most frequent pair
 end while
 erase rules below freq_accept
 cpx= complexity of sequence(s)
 if ( cpx <= cp_min) then
     cpx_min= cpx

  Complexity of sequence(s) is cpx_min

 end for
```

**Fig 2: Pseudocode**